# Orbital Angular Momentum (OAM) Carrying Vortex Wave generation in Dielectric Filled Circular Waveguide

Md Khadimul Islam, Arjuna Madanayake and Shubhendu Bhardwaj
Department of Electrical and Computer Engineering, Florida International University, Miami, FL, USA
Email: {misla081, amadanay, sbhardwa}@fiu.edu

*Abstract*—In this paper, we propose a method to generate Orbital Angular Momentum (OAM) carrying vortex waves inside a metallic circular waveguide (CW). These waves feature ability to carry multiple orthogonal modes at the same frequency, by the virtue of their unique spatial structure. In essence high data rate channels can be developed using such waves. In free space, OAM carrying vortex waves has beam divergence issues and a central NULL, which makes the waves unfavourable for free space communication. But, OAM modes in guided structures do not suffer from these drawbacks. This prospect of enhancement of communication spectrum provides the background for the study of vortex wave in the circular waveguides. In this work, a radial array of monopoles is designed to generate the vortex wave inside the waveguide. Further, we introduced the dielectric materials inside the waveguide in order to manipulate the operating frequency of the OAM modes. Simulation results shows that the various dielectric materials allow us to tune the working frequency of the OAM beam to a desired frequency.

## I. INTRODUCTION

A propagating wave with vortex behavior in the phase front following $e^{jl\varphi}$ can carry the Orbital Angular Momentum (OAM), where $\varphi$ is azimuthal angle and $l$ is the topological charge, commonly known as OAM mode number [1]. With the advancement of the wireless communication technology, the data-rate requirement is continuously increasing. The current state of art data-multiplexing techniques are not sufficient to furnish the continuously increasing data-rate requirement. Vortex waves offers a potential solution to that issue in the wireless spectrum by providing infinite number of channels based on the twisting angle of the phase-front [2] [3]. Therefore, there is an increasing interest in the multimodal vortex wave communication.

Any electromagnetic system produces Angular Momentum (AM) beside it's linear momentum. The AM consists of Spin Angular Momentum (SAM) which is related to circular polarization and Orbital Angular Momentum (OAM) which is associated to the spatial distribution of the phase. Notably, as per prior experiments in optics, many waves with varying orbital angular momentum can be multiplexed in a region of space and at the same frequency [4] [5]. The similar outcomes are obtained at the microwave frequencies, which provides the enhancement of the wireless spectrum. Several successful attempts have been made to demonstrate OAM beams' generation, including the use of spiral phase plate (SPP), linear phase plate (LPP), reflector antennas, circular array and so on [6] [7] [5]. Those structure are used to create the required phase-twist by introducing an azimuth-dependent path delay to the wave. In some instances, a superposition technique, featuring an addition of multiple fundamental modes with appropriate phase delays, is used to produce the vorticity [4]. In this paper, the generation of the vortex wave modes is demonstrated using full wave simulation inside the metallic circular waveguide using a radial array of monopoles. The proper phase assignments resemble the field profile of the two targeted fundamental waveguide modes and generate the vorticity on the phase front. Furthermore, the manipulation of the operating frequency of the OAM modes by introducing the dielectric materials inside waveguide is also demonstrated.

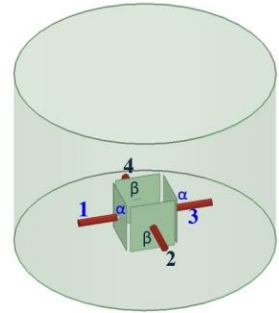

Fig. 1: The circular waveguide with radial monopole array.

## II. OAM MODES INSIDE AN IDEAL WAVEGUIDE

It is evident from the solution of orthogonal TE eigenmodes of a circular waveguide that the TE modes are the combination of two degenerate modes denoted by $TE_{clm}$ and $TE_{slm}$, which show the same instantaneous E-field distribution with $\pi/4$ phase difference [4]. The superposition of those two modes with $\pi/2$ phase will result in the rotating phase front, i.e. the OAM carrying vortex wave will be generated. Hence the combination of the modes will be as follows,

$$TE_{clm} + iTE_{slm} \qquad (1)$$

To test the idea, we designed a metallic circular waveguide of radius 2 cm and length 3 cm, with waveguide cut-off at 4.4 GHz (see Fig. 1). The combination of the two degenerate modes $TE_{21}$ and $TE_{41}$ is used to subsequently create the OAM $l = 1$ and $l = 3$ modes. Fig. 2 shows the magnitude and phase of the $E_x$ component. It is seen from the phase

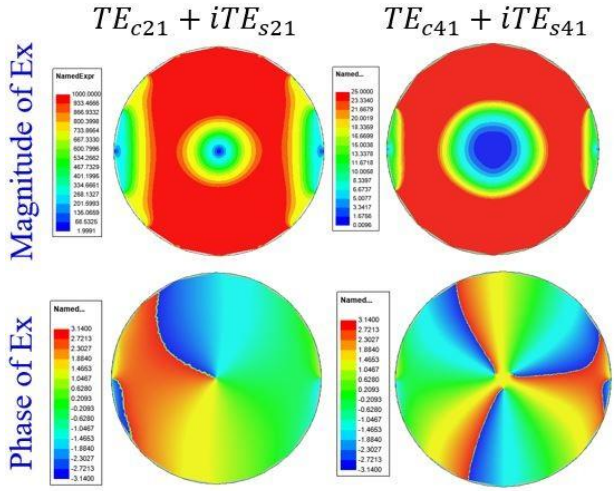

Fig. 2: OAM $l = 1$ and $l = 3$ mode generation inside ideal waveguide.

plot that the phase undergoes a change from $-\pi$ to $\pi$ times it's OAM mode number

### III. OAM Mode Generation using Proposed Antenna Structure

The generation of OAM modes is shown by super-positioning of the fundamental modes is discussed above. To implement this, we propose a structure consisting of a radial array of four monopoles as shown in Fig. 1. The lengths of the antennas are considered to be 6 mm to provide resonance at around 10 GHz. The antennas 1 3 (see Fig.

1) are excited with the same phase $\alpha$ and antenna 2 4 are excited with the same phase $\beta$. First we mimic the ideal electric field due to two $TE_{21}$ degenerate modes inside the waveguide using the proposed antenna structure, considering $\alpha = \beta = 0$. Then we applied the $\pi/2$ phase difference between two sets of antenna to generate the vortex wave in Fig. 3, which shows the obtained vortex wave from the structure with $\alpha = 0$ and $\beta = \pi/2$.

We also introduced a dielectric material inside the waveguide for frequency tuning. Presence of 8 mm thick dielectric material with $\varepsilon_r = 2.5$ changed the cut-off frequency of $TE_{21}$ mode of the waveguide from 7.2 GHz to 5.8 GHz and the antennas see a different radiation space. As a result, the same structure is generating the OAM $l = 1$ mode in 6.5 GHz in stead of 10 GHz. Fig. 4 shows the structure with dielectric inside and the generated vortex wave at 6.5 GHz.

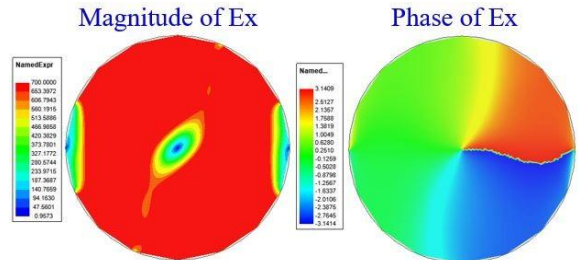

Fig. 3: Generation of OAM $l = 1$ mode using the proposed structure at the frequency of 10 GHz.

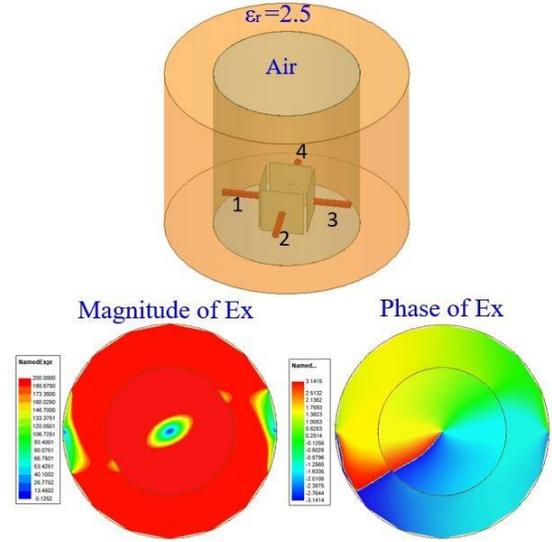

Fig. 4: Generation of OAM $l = 1$ mode at 6.5 GHz using the proposed structure with dielectric inside.

### IV. Conclusion

We proposed and showed the generation of OAM carrying vortex wave using radial array of four monopoles. Using the proposed structure, first an ideal E-field is generated and proper phase difference is applied between the antenna sets to create the vortex wave. Further, the shift in the operating frequency is observed due to the introduction of dielectric materials which indicates that the tuning of OAM modes to any desired frequency is possible. Following in the similar manner, higher order OAM modes at any desired frequency can be generated using the multiple antenna set with proper phase assignment.